\documentclass[apjl]{emulateapj}

\usepackage[usenames]{color}
\usepackage{comment}

\newcommand{\fourstar}{{\sc FourStar}}
\newcommand{\jo}{$J_1$}
\newcommand{\jt}{$J_2$}
\newcommand{\jh}{$J_3$}
\newcommand{\hs}{$H_s$}
\newcommand{\hl}{$H_l$}
\newcommand{\ks}{$K_s$}
\newcommand{\mh}{$M_{halo}$}
\newcommand{\ms}{$M_{\ast}$}
\newcommand{\msun}{M$_{\odot}$}

\slugcomment{accepted by ApJL}
\shorttitle{A $z=2.2$ candidate galaxy cluster}
\shortauthors{Spitler et al.}

\begin{document}
\title{First results from $Z-$FOURGE\altaffilmark{$\star$}:  Discovery of a Candidate Cluster at $z=2.2$ in COSMOS}\notetoeditor{Can the \altaffilmark{$\star$} be appended to Z-FOURGE in the title?}
\email{lspitler@astro.swin.edu.au}

\author{Lee~R.~Spitler\altaffilmark{1},
Ivo~Labb\'e\altaffilmark{2,}\altaffilmark{3},
Karl~Glazebrook\altaffilmark{1},
S.~Eric~Persson\altaffilmark{2},
Andy~Monson\altaffilmark{2},
Casey~Papovich\altaffilmark{4},
Kim-Vy~H.~Tran\altaffilmark{4},
Gregory~B.~Poole\altaffilmark{1},
Ryan~Quadri\altaffilmark{2,}\altaffilmark{7},
Pieter~van~Dokkum\altaffilmark{5},
Daniel~D.~Kelson\altaffilmark{2},
Glenn~G.~Kacprzak\altaffilmark{1,}\altaffilmark{6},
Patrick~J.~McCarthy\altaffilmark{2},
David~Murphy\altaffilmark{2},
Caroline~M.~S.~Straatman\altaffilmark{3},
Vithal~Tilvi\altaffilmark{4}
}

\altaffiltext{1}{Centre for Astrophysics \& Supercomputing, Swinburne University, Hawthorn, VIC 3122, Australia}
\altaffiltext{2}{Carnegie Observatories, Pasadena, CA 91101, USA}
\altaffiltext{3}{Sterrewacht Leiden, Leiden University, NL-2300 RA Leiden, The Netherlands}
\altaffiltext{4}{George P.\ and Cynthia Woods Mitchell Institute for Fundamental  Physics and Astronomy, and Department of Physics and Astronomy, Texas A\&M University, College Station, TX, 77843-4242, USA}
\altaffiltext{5}{Department of Astronomy, Yale University, New Haven, CT 06520, USA}
\altaffiltext{6}{Australian Research Council Super Science Fellow}
\altaffiltext{7}{Hubble Fellow}
\altaffiltext{$\star$}{This paper includes data gathered with the 6.5 meter Magellan Telescopes located at Las Campanas Observatory, Chile.}

\begin{abstract}
We report the first results from the $Z-$FOURGE survey: the discovery of a candidate galaxy cluster at $z=2.2$ consisting of two compact overdensities with red galaxies detected at $\ga20\sigma$ above the mean surface density.  The discovery was made possible by a new deep ($K_s\la24.8$ AB $5\sigma$) Magellan/\fourstar\ near-IR imaging survey with 5 custom medium-bandwidth filters. The filters pinpoint the location of the Balmer/4000\AA\ break in evolved stellar populations at $1.5<z<3.5$, yielding significantly more accurate photometric redshifts than possible with broadband imaging alone.  The overdensities are within $1\arcmin$ of each other in the COSMOS field and appear to be embedded in a larger structure that contains at least one additional overdensity ($\sim10\sigma$).  Considering the global properties of the overdensities, the $z=2.2$ system appears to be the most distant example of a galaxy cluster with a population of red galaxies.  A comparison to a large $\Lambda$CDM simulation suggests that the system may consist of merging subclusters, with properties in between those of $z>2$ protoclusters with more diffuse distributions of blue galaxies and the lower-redshift galaxy clusters with prominent red sequences.  The structure is completely absent in public optical catalogs in COSMOS and only weakly visible in a shallower near-IR survey.  The discovery showcases the potential of deep near-IR surveys with medium-band filters to advance the understanding of environment and galaxy evolution at $z>1.5$.
\end{abstract}

\keywords{galaxies: high-redshift --- galaxies: clusters: general --- large-scale structure of Universe}

\section{Introduction}\label{intro}

Galaxy clusters are the most overdense cosmological regions in the Universe and are unique astronomical tools: their abundance constrains fundamental cosmological parameters and they provide an extreme laboratory for elucidating the role of local environment in the evolution of the massive galaxies.

However, despite extensive multi-wavelength searches, only a few evolved galaxy clusters with red galaxies have been found at $z\ga1.5$ \citep[e.g][]{mccarthy_compact_2007,andreon_jkcs_2009,papovich_spitzer-selected_2010,tanaka_spectroscopically_2010,fassbender_x-ray_2011,gobat_mature_2011,santos_discovery_2011}.  More diffuse protoclusters of star-forming galaxies have been found to higher redshift \citep[e.g.][]{steidel_ly_2000,venemans_protoclusters_2007,capak_massive_2011}, although it is unclear how they relate to lower redshift massive clusters.

Ideally we want to find massive distant structures using spectroscopy. The problem is that most cluster galaxies at $z>1.5$ are too faint for spectroscopy, while photometric redshifts derived from broadband photometry are generally not accurate enough for secure identification.  A novel approach is to use near-infrared imaging with {\it medium-bandwidth} filters, which are narrower than traditional broadband filters and provide significantly more accurate photometric redshifts \citep[e.g., $\delta z/(1+z)\sim1-2\%$ at $z\sim2$;][]{van_dokkum_newfirm_2009,whitaker_newfirm_2011} for thousands of sources simultaneously over large contiguous fields of view \citep[e.g., ][]{wolf_combo-17_2003}.

Using the newly commissioned \fourstar\ near-IR camera \citep{persson_fourstar_2008} on the 6.5m Magellan Baade Telescope, we have initiated a major survey to obtain deep medium-bandwidth near-IR imaging over several fields. Our custom filters span $1.0\micron-1.8\micron$ and hence trace the Balmer/4000\AA\ break in galaxies at $1.5<z<3.5$.  The \fourstar\ Galaxy Evolution Survey ($Z-$FOURGE\footnote{{\url http://z-fourge.strw.leidenuniv.nl/}}) will be described in detail by Labb\'e et al.,(in prep).  This Letter demonstrates that the improved depth and redshift accuracies allow us to search for massive galaxy overdensities at redshifts $z\ga1.5$.

\section{Observations and analysis}

\begin{figure*}
\centering
\includegraphics[width=\textwidth]{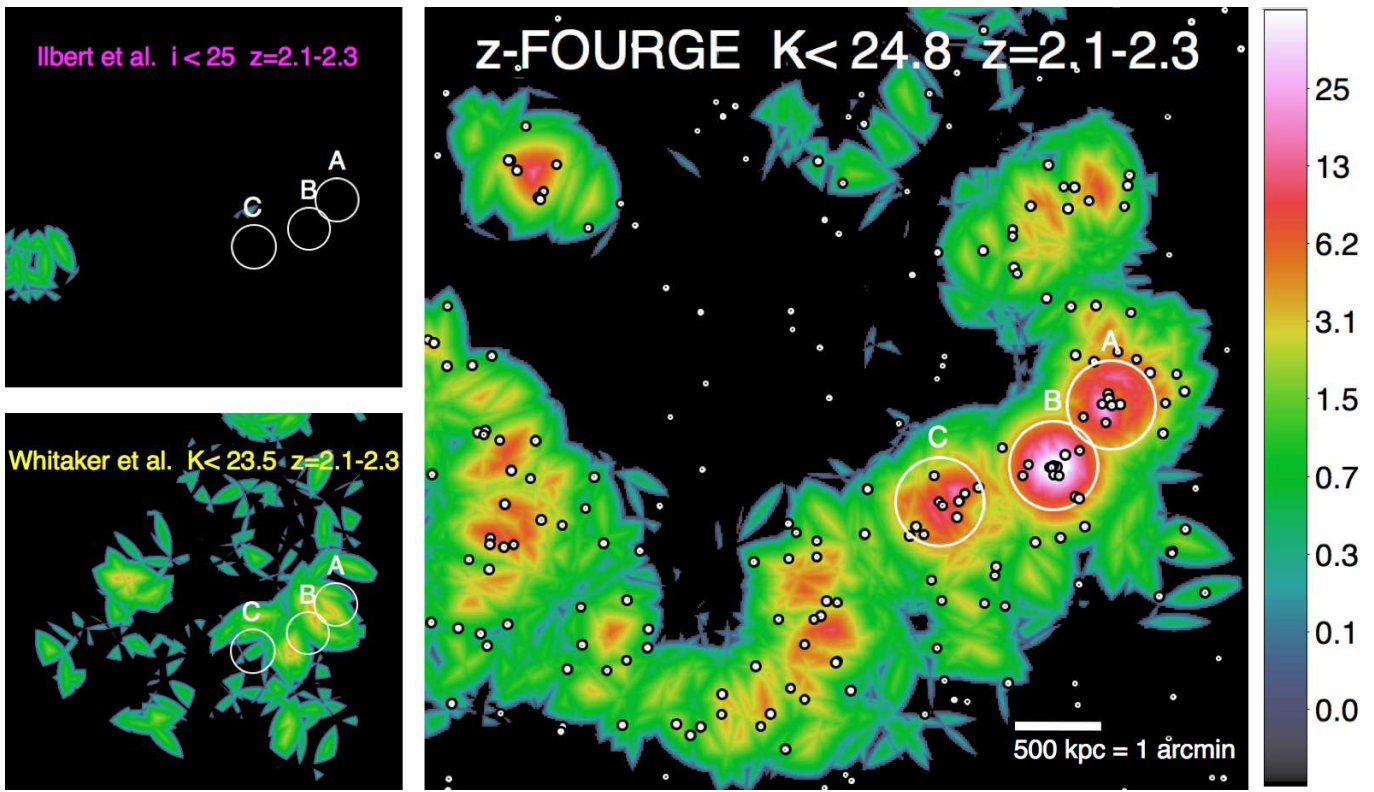}
\caption{7th~nearest-neighbor surface density maps for $z=2.1-2.3$ in a $\approx9\arcmin\times9\arcmin$ region in the COSMOS field. Units are standard deviations above the mean. Density maps, including those from literature photometric redshift catalogs \citep{ilbert_cosmos_2009,whitaker_newfirm_2011}, are labeled along with the limiting selection magnitude.  Individual $Z-$FOURGE galaxies at $z=2.1-2.3$ are represented by small circles.  The maps illustrate the advantage of deep near-infrared imaging with medium-band filters for finding large-scale structures at $z\sim2$.\label{fig_density}}
\end{figure*}

\notetoeditor{Put Figure~\ref{fig_density} at top of 2nd page if possible.}

As part of the on-going survey we observed a single $\approx11\arcmin\times11\arcmin$ pointing within the COSMOS field \citep{scoville_cosmic_2007} with Magellan/\fourstar\ in Spring 2011, imaging for 41 hours in 5 medium-bandwidth filters (\jo, \jt, \jh, \hs, \hl) and the broad-band \ks\ filter to $5\sigma$ point-source limiting depths (D$=1.5\arcsec$ aperture corrected to total) of 25.5, 25.4, 25.3, 24.8, 24.7 and 24.8 AB mag, respectively. Raw images were processed using our custom pipeline, also used for the NEWFIRM Medium Band Survey \citep[NMBS,][]{whitaker_newfirm_2011}.

{\tt SExtractor} \citep{bertin_sextractor:_1996} was used to select objects in the $K_s-$image (FWHM $\approx0.40\arcsec$; 0.15\arcsec pixel$^{-1}$) to a depth of $K_s\sim24.5$ and to extract D$=1.5\arcsec$ aperture fluxes from PSF-matched versions of our $Z-$FOURGE COSMOS data, plus 23 optical and 4 Spitzer/IRAC COSMOS legacy image sets. Photometric zeropoints were calibrated using sources in common with the NMBS catalogue in COSMOS \citep{whitaker_newfirm_2011}.  We adopt the AB magnitude system and a cosmology with $H_0 = 70$ km s$^{-1}$ Mpc$^{-1}$, $\Omega_m=0.3$, and $\Omega_{\Lambda}=0.7$.  Stars were culled using a $U-J_1$ and $J_1-K_s$ color-color criterion \citep{whitaker_newfirm_2011}.

We derived photometric redshifts with {\tt EAZY} \citep{brammer_eazy:_2008} and fitted \citet{bruzual_stellar_2003} stellar population models using {\tt FAST} \citep{kriek_ultra-deep_2009}, assuming exponentially declining star-formation histories, Solar metallicity and a \citet{chabrier_galactic_2003} initial mass function.  We note that photometric redshifts derived with similar near-IR medium-band filters yielded a normalized median absolute deviation of $\delta_{z,nmad}/(1+z)=2$\% at $z=1.7-2.7$ in the NMBS survey \citep{van_dokkum_newfirm_2009}.

\section{Overdensities at $z=2.2$}

\begin{figure*}
\centering
\includegraphics[width=\textwidth]{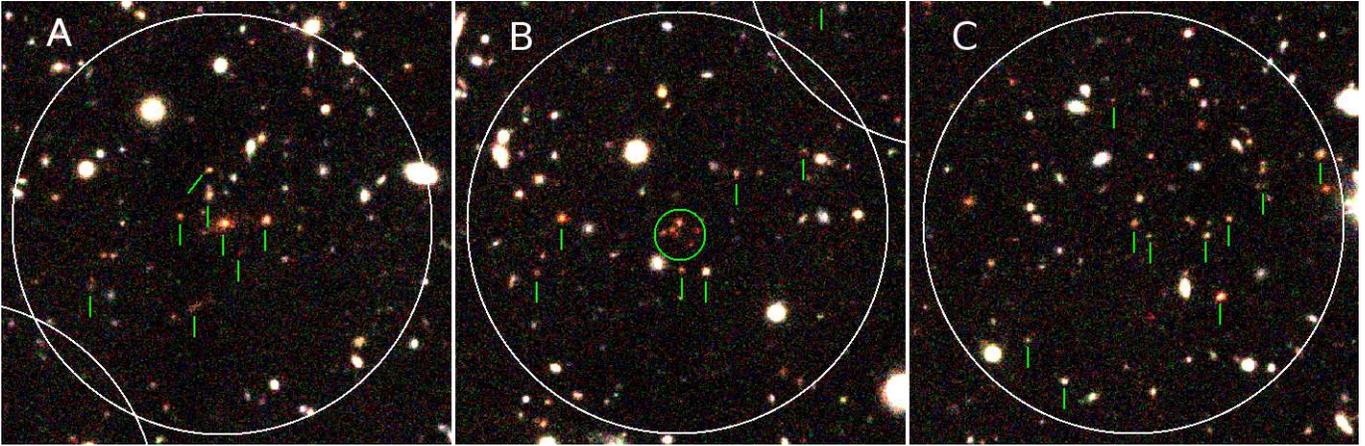}
\caption{ Color composite images in the \jo~(blue,$1.05\micron$), \jt~(green,$1.15\micron$) and \jh~(red,$1.28\micron$) $\Delta\lambda/\lambda\sim10\%$ filters, centered on each overdensity in the $Z-$FOURGE observations with \fourstar\ in the COSMOS field. Galaxies at $z=2.0-2.3$ with strong Balmer/4000\AA\ breaks have red colors in this image and are marked with green symbols.  White circles have $r=30\arcsec$. \label{fig_composite}}
\end{figure*}

\subsection{Discovery}\label{discovery}

We searched for high-redshift galaxy overdensities by computing surface density maps in narrow $\delta z=0.2$ redshift slices between $z=1.5-3.5$ using the 7th nearest-neighbor metric \citep[e.g., ][]{papovich_spitzer-selected_2010,gobat_mature_2011}.  At each location in the map, we calculate the projected distance to the 7th nearest neighbor and evaluate the projected density $n_{7}=N/(\pi*r_N^2)$.  The results do not change significantly for density maps with $N=5-9$.

As shown in Fig.~\ref{fig_density}, a system consisting of 3 strong galaxy overdensities within a radius of 1.5\arcmin\ was found between $z=2.1-2.3$.  At these redshifts the Balmer/4000\AA\ break passes through the \fourstar\ $J_1(1.05\micron)$, $J_2(1.15\micron)$, $J_3(1.28\micron)$ medium-band filters.  Fig.~\ref{fig_composite} shows images of the overdensities, which each contain dense concentrations of galaxies with red $J_2-J_3$ colors, consistent with the presence of prominent Balmer/4000\AA\ breaks at $z\sim2.2$.

\begin{figure}
\centering
\includegraphics[angle=-90,scale=.35]{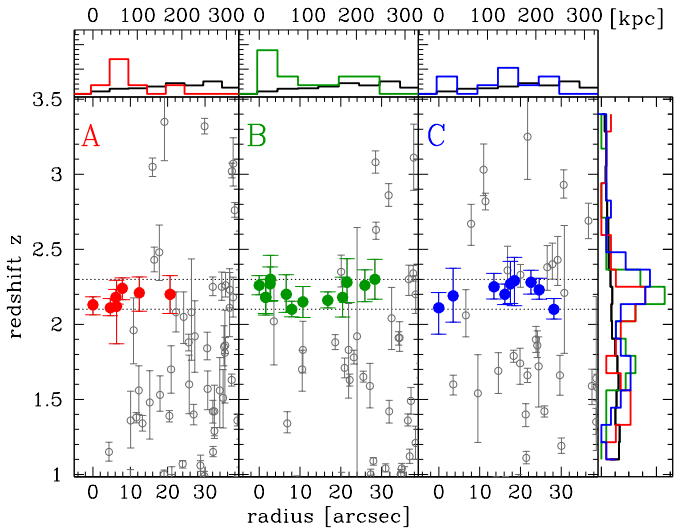}
\caption{Photometric redshift against radial distance to the central brightest galaxy in the labeled overdensities ABC from Fig.~\ref{fig_density}. Colored points and histograms are galaxies with $r<30\arcsec$ and $2.1\le z\le2.3$. The histograms are normalized by the area from which the sample was drawn.  The control histograms (black) in the top panels are the cumulative spatial distribution around all $z=2.1-2.3$ galaxies. The control histograms in the right panel are all galaxies in the field.  The overdensities show concentrated surface densities and peaked redshift distributions compared to the control samples. 
\label{fig_redshift}}
\end{figure}

\begin{figure}
\epsscale{1.2}
\plotone{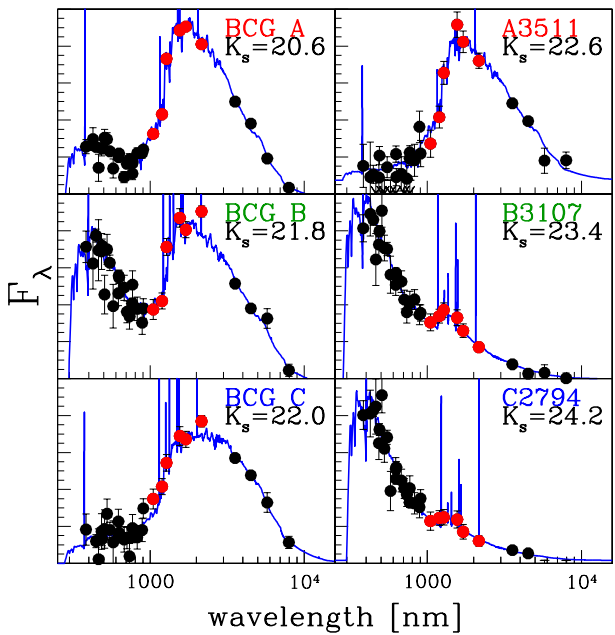}
\caption{Observed 32-band SEDs of the central brightest galaxies (BCGs) in the overdensities (left panels) and randomly selected overdensity members (right panels). The highlighted red points are our custom \fourstar\ filters, sensitive to near-IR light at $1.0-1.8$ micron. The higher resolution spectral sampling of the SEDs in this range allow us to pinpoint the Balmer/4000\AA\ break as it shifts through medium-bandwidths $1.5<z<3.5$.  Overplotted are the best-fit {\tt EAZY} photometric redshift templates (blue).  BCG~A has a quiescent stellar population, while BCGs B \& C have some star formation, including enhanced $K_s-$band flux likely due to H$\alpha$ emission.
\label{fig_sed}}
\end{figure}

Fig.~\ref{fig_redshift} presents photometric redshifts for the overdensity galaxies and further confirms that they are strongly peaked at $z\approx2.2$. The overdensities have consistent mean redshifts ($z^A=2.16\pm0.03$, $z^B=2.19\pm0.03$, $z^C=2.21\pm0.03$; uncertainties are random error on the mean) and together show a weighted mean of $2.19\pm0.03$ (here we adopt as the uncertainty the range of the 3 overdensity redshifts).  The RMS scatter of individual galaxies is $0.06$ or $\delta z/(1+z)=0.02$.  There are 7, 13, \& 9 candidate members within $<30\arcsec$ of the overdensities\footnote{We adopt the brightest galaxy as an overdensity's center:  (10:00:15.753, +02:15:39.56), (10:00:18.380, +02:14:58.81), and (10:00:23.552, +02:14:34.13), for overdensity A, B \& C, respectively (J2000).} A, B, C, respectively. The red objects furthermore have steep Balmer/4000\AA\ breaks between the $J_2$ and $J_3$ filters, as is shown in Fig.~\ref{fig_sed}.  

We also calculated a density map using existing photometric redshift catalogs in COSMOS \citep{ilbert_cosmos_2009,whitaker_newfirm_2011} in Fig.~\ref{fig_density}.  The overdensities are completely absent in the public $i-$band selected catalog of \citet{ilbert_cosmos_2009}. Only a weak impression of the overdensities is apparent in the $K_s-$band selected catalog of \citet{whitaker_newfirm_2011}.  This substantiates the critical role that deep, near-IR imaging with medium-band filters will play in understanding environment and galaxy evolution at redshifts $z>1.5$.  

We note the candidate cluster satisfies the Spitzer/IRAC color based selection criteria of \citet{papovich_angular_2008}, used to discover a $z=1.62$ cluster \citep{papovich_spitzer-selected_2010}.  However, unlike the IRAC selection, our catalogs provide accurate photometric redshifts, thus reducing spurious detections from foreground interlopers and enabling the secure identification of an overdensity at $z\sim2.2$.

\subsection{Significance of the overdensities}\label{overdensity_sig}

To quantify the statistical significance of the overdensities, we first estimated the mean and intrinsic scatter in the nearest neighbor density map of Fig.~\ref{fig_density}. To avoid biasing these values by the strong overdensities themselves, we use the mean density ($n_7=2.6$ arcmin$^{-2}$) and its standard deviation ($\sigma_{n7}=1.4$ arcmin$^{-2}$) from adjacent redshift slices ($z=1.9-2.1$ and $z=2.3-2.5$). These statistics reflect the distribution of nearest neighbor densities evaluated only at the locations of all galaxies in a redshift slice. We find that the overdensities are $\approx20$, $50$, and $10\sigma$ deviations for A, B, and C, respectively.

We also performed a bootstrap resampling of the \fourstar\ redshifts.  At each instance, we shuffled all redshifts in our catalog and generated a 7th nearest-neighbor density map.  To robustly identify overdensities in the resampled maps, we tuned {\tt SExtractor} ({\tt DETECT\_THRESH}, {\tt SEEING\_FWHM}) to detect only overdensities A \& B in the real density map.  In only 3 of the 1000 resampled maps was {\it a single} overdensity detected.  When tuned to find the less significant overdensity~C, {\tt SExtractor} detects only 65 overdensities in the resampled maps.  Note the number of valid analogs in the resampled maps would decrease further if we tried to match the tight spatial configuration of the real overdensities.

As a final check, we analyzed 121 mock density maps from simulated light cones produced by the Mock Galaxy Factory (Bernyk et al.,in prep.). These are based upon the Millennium Simulation \citep{springel_simulations_2005} and semi-analytical models of \citet{croton_many_2006}.  After introducing fake redshift errors, we matched the number of observed galaxies in the $Z-$FOURGE COSMOS field by selecting an $R$-band absolute magnitude limit $M_R<-21.6$ (roughly $K_s\la24.5$ at $z=2.2$) and found a consistent scatter ($\sigma_{n7}=2.0\pm0.7$ arcmin$^{-2}$) with our own estimate.

The above results confirm that overdensities A \& B are robust, while overdensity C appears to be slightly less significant.  Its close proximity to A \& B raises the intriguing possibility that it is associated with the AB system. We therefore include overdensity~C in the following.

\section{Candidate cluster properties}

\subsection{Galaxy properties}\label{stellarpops}

\begin{figure*}
\epsscale{0.5}
\plotone{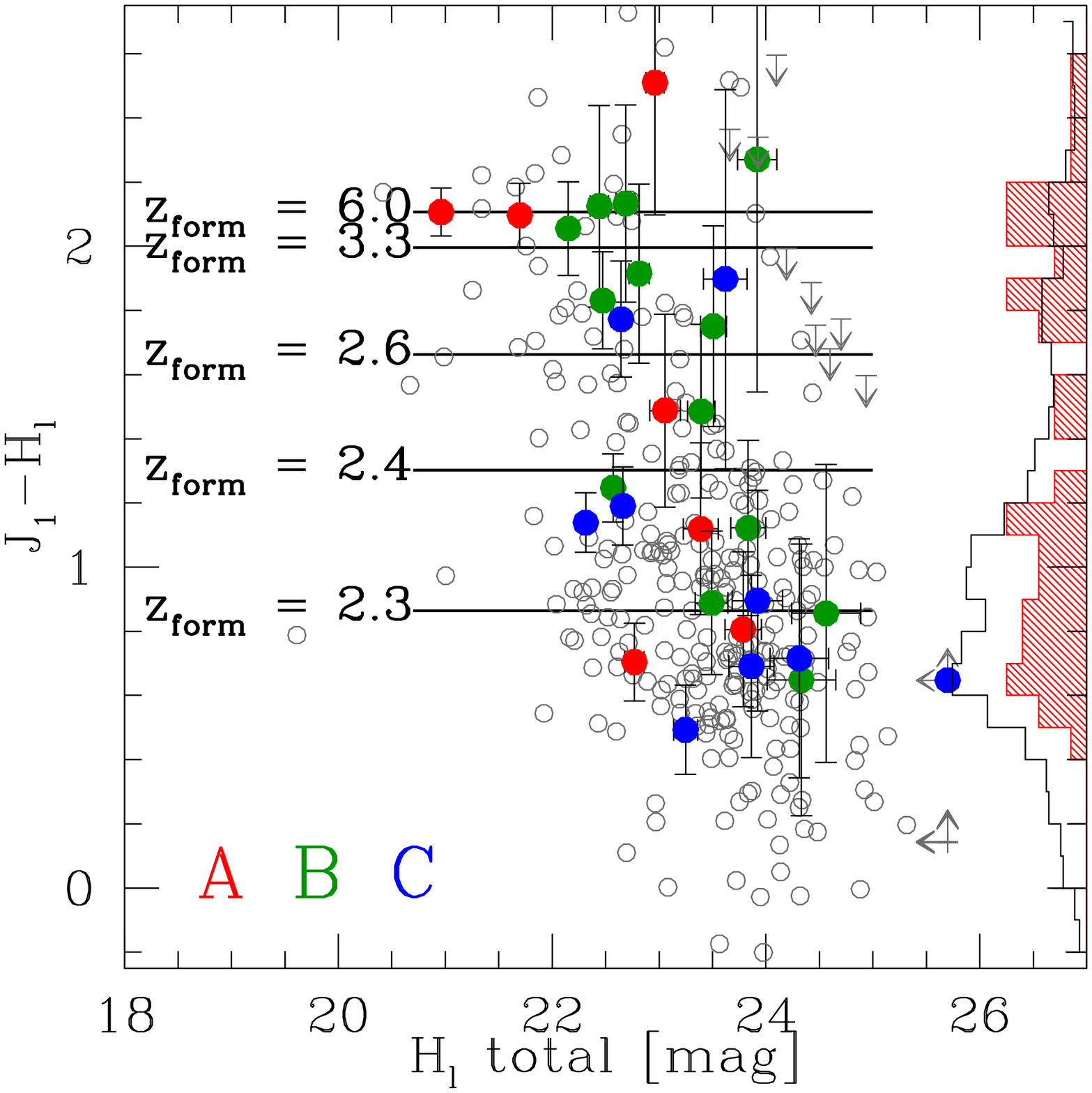}
\plotone{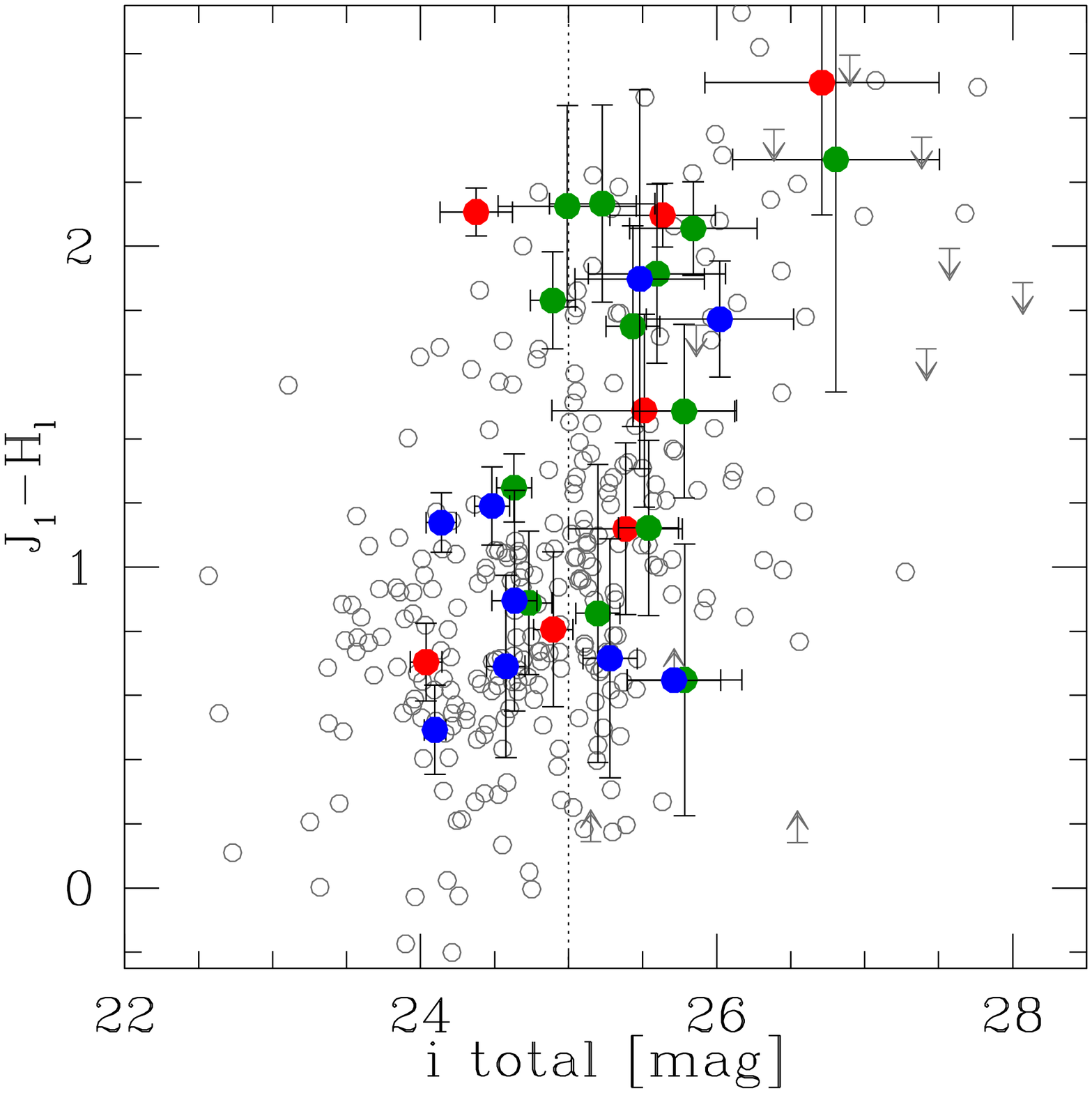}
\caption{$J_1-H_l$ versus $H_l$ ({\it left panel}) and versus $i$ ({\it right panel}) color-magnitude diagrams. Grey points and open histograms are for all 313 galaxies with $2.1\le z\le2.3$.  Galaxy overdensity members are shown with the colored points and red hatched histograms.  Histograms are arbitrarily normalized. Limits correspond to $2\sigma$ near-IR limiting depths.  The overdensities have a higher fraction of red galaxies compared to the field. In the left panel, black lines show the $z=2.2$ redshifted \citet{bruzual_stellar_2003} single-burst model predictions with Solar metallicity and a \citet{chabrier_galactic_2003} initial mass function for various ages or formation redshifts ($z_{form}$; computed with {\tt EzGal}, {\tt www.baryons.org/ezgal}). In the right panel, red galaxies are generally very faint in the optical $i-$band. E.g., 80\% of galaxies redder than $J_1-H_l>1.6$ are fainter than $i=25$ (dotted line).\label{fig_cmd}}
\end{figure*}

Of the 313 galaxies in the redshift slice $2.1\le z\le2.3$ over the full $Z-$FOURGE COSMOS field, 29 galaxies are within $30\arcsec$ of a $z=2.2$ overdensity. We consider these candidate overdensity galaxies.

Fig.~\ref{fig_cmd} shows observed color-magnitude diagrams for all galaxies having $2.1\le z\le2.3$.  The $J_1-H_l$ color probes continuum on both sides of the 4000\AA\ break and avoids rest-frame H$\alpha$ in \ks.  The histograms in Fig.~\ref{fig_cmd} show that the non-overdensity or ``field'' distribution is dominated by blue galaxies ($J_1-H_l<1.6$) while the overdensity galaxies have a higher fraction of red galaxies.

We calculated the red galaxy fractions, $f_{red}=N_{red}/N_{total}$, of each overdensity.  Here $N_{red}$ is the number of galaxies with $J_1-H_l\ge1.6$ at {\it all} magnitudes within $r<30\arcsec$ of an overdensity at $2.1\le z\le2.3$.  Apart from C ($f_{red}=0.2\pm0.2$), the two main overdensities show somewhat higher red galaxy fractions (together $f_{red}=0.5\pm0.2$) compared to the field population ($f_{red}=0.20\pm0.03$; $f_{red}$ errors reflect counting statistics only).

Fig.~\ref{fig_cmd} shows that the red galaxies have similar colors to BC03 single-burst stellar population models at $z=2.2$ and formation redshifts of $z_{form}\ga3$.  Fitting BC03 models with exponentially declining star formation histories to the full SEDs confirms that the red galaxies are on average $\sim1$ Gyr old ($z_{form}\sim3.3$), contain little on-going star formation and span a stellar mass range of \ms$=0.1-5\times10^{11}M_{\odot}$. In contrast, blue overdensity galaxies ($J_1-H_l<1.6$) are best-fit by $\sim0.1$ Gyr BC03 models and have masses \ms$=10^8-10^{10}M_{\odot}$.

We find candidate ``brightest cluster galaxies'' (BCGs) in each overdensity, with relatively large stellar masses:  $M^A_{\ast}=3\times10^{11}$, $M^B_{\ast}=1\times10^{11}$, $M^C_{\ast}=1\times10^{11}$\msun.  As shown in Fig.~\ref{fig_sed}, BCG~A is a quiescent ``red and dead'' galaxy, while BCGs B \& C may contain recent star formation.  A close inspection of BCGs A \& B in Fig.~\ref{fig_composite} suggests they also have distinct structural properties:  the latter is compact\footnote{As measured on the \ks-band \fourstar\ image using 2D S\'ersic models fitted with {\tt ISHAPE}, which is specifically designed to measure sizes of partially resolved objects \citep{larsen_young_1999}.} ($r_e=2^{+0.3}_{-0.5}$kpc) while the former has a larger core ($r_e=3^{+0.5}_{-0.5}$~kpc) plus an extended diffuse stellar halo.  With $3-4$ satellites within $r\sim30$~kpc, we speculate BCG~B will undergo a series of mergers and perhaps ``puff-up'' in size \citep{hopkins_discriminating_2010}.  Clearly we are probing an epoch where even some of the most massive galaxies in the highest density regions were still forming a significant fraction of their stars \citep[e.g., ][]{glazebrook_high_2004,van_dokkum_star_2007,eisenhardt_clusters_2008,tran_reversal_2010}.

\subsection{Comparison to known high-$z$ overdensities}\label{highz}

To help us interpret the $z=2.2$ overdensities, we will now characterize various global properties of the overdensities and compare them to known high-redshift galaxy clusters and protoclusters.

As shown in Fig.~\ref{fig_density}, the individual $z=2.2$ overdensities have spatial extents of $r=30\arcsec$ or $250$ kpc. The galaxy clusters at $z\ga1.6$ show similar projected compact sizes \citep{andreon_jkcs_2009,papovich_spitzer-selected_2010,tanaka_spectroscopically_2010,gobat_mature_2011}.  Notably, the $z=1.62$ cluster \citep{papovich_spitzer-selected_2010,tanaka_spectroscopically_2010} shows two galaxy clumps or subclusters over a region of $\sim1$\arcmin. This is not unlike the configuration discussed here. In contrast, known protoclusters at $z\ga2$ are typically more diffuse, with lower overdensities and $\sim8\times$ larger sizes \citep{steidel_ly_2000,venemans_protoclusters_2007}.  Indeed, overdensities A \& B each show core surface densities of $\ga50$ galaxies arcmin$^{-2}$, with $5-6$ members at $r\la10\arcsec$.

To estimate the total halo mass (\mh) of each overdensity, we use the relation between \ms\ and \mh\ at $z=2.2$ from the halo occupancy distribution analysis of \citet{moster_constraints_2010} and apply these to the \ms\ of the central BCGs.  Although the uncertainties involved in converting stellar mass to halo mass are significant, we estimate that the overdensities have: $M_{halo}^A\approx6\times10^{13}$, $M_{halo}^B\approx1\times10^{13}$, $M_{halo}^C\approx1\times10^{13}$\msun. The \mh\ of A is in the same range as estimates for the $z=2.07$ cluster \citep{gobat_mature_2011}.  Using a 1 Gpc$^3$ cosmological simulation (GiggleZ; Poole et al.,in prep.; $2160^3$ particles, WMAP5 cosmology, \citealt{komatsu_five-year_2009}), we find that $z=2.2$ simulated halos with these masses will grow into $z=0$ halos with mean masses of $M_{halo}\sim0.5-5\times10^{14}$\msun.

The detection of diffuse X-ray emission would provide independent confirmation of the existence of a deep gravitational potential well. We find no diffuse X-ray emission at the locations of the $z=2.2$ overdensities in the COSMOS 55ks XMM and 200ks Chandra legacy images \citep{hasinger_xmm-newton_2007,elvis_chandra_2009,cappelluti_xmm-newton_2009} to a point-source $0.5-2$ keV sensitivity limit of $\sim2\times10^{-15}$ and $2\times10^{-16}$ erg cm$^{-2}$~s$^{-1}$, respectively.  The XMM flux limit corresponds to an X-ray luminosity upper limit of $7\times10^{43}$ erg~s$^{-1}$, similar to the value estimated for the extended emission around the $z=2.07$ \citet{gobat_mature_2011} cluster.

Overall, the newly discovered $z=2.2$ overdensities show a number of characteristics (e.g., \mh\ estimates, presence of evolved massive galaxies, compact spatial distribution) similar to spectroscopically-confirmed galaxy clusters at $1.5<z<2.1$ \citep{papovich_spitzer-selected_2010,tanaka_spectroscopically_2010,gobat_mature_2011}.  Even so, there are also marked variations: e.g., overdensity~C contains mostly blue star-forming galaxies, more resembling our field population at $z=2.1-2.3$. Given the overall resemblance and the close proximity of the three overdensities on the sky, we consider it likely that all three overdensities may be part of a single forming massive cluster.

As shown in Fig.~\ref{fig_density}, the region surrounding the $z=2.2$ system contains several less-significant galaxy overdensities, as one might expect to find around a large overdensity \citep[e.g., ][]{springel_large-scale_2006}.  The overdensity just to the north of the $z=2.2$ system is particularly interesting:  it contains two \ms$\approx1-5\times10^{11}$\msun\ evolved galaxies separated by just $20\arcsec$.

\section{Discussion and Conclusions}\label{discussion}

Using the first data from the Magellan/\fourstar\ Galaxy Evolution Survey ($Z-$FOURGE), we have detected in the COSMOS field the most distant example of strong galaxy overdensities with red galaxies.  The system at $z=2.2$ extends $r\sim1.5\arcmin$ on the sky and appears to be made up of multiple subcomponents, including two overdensities detected at $\ga20\sigma$ above the mean and another at $\sim10\sigma$.  The two strongest overdensities, A \& B, each resemble spectroscopically-confirmed galaxy clusters at $z\ga1.6$ \citep{papovich_spitzer-selected_2010,tanaka_spectroscopically_2010,gobat_mature_2011}: they contain significant populations of evolved galaxies (mean stellar age $\sim1$~Gyr or $z_{form}\sim3.3$) in a compact spatial distribution ($<250$kpc) embedded in massive halos of \mh$\sim1-6\times10^{13}$\msun.

Perhaps the most outstanding feature of the aggregate system is that it consists of multiple distinct galaxy overdensities in close spatial configuration.  Whether the three $z=2.2$ overdensities are a single gravitationally bound structure is not clear.  Ultimately, high precision spectroscopic redshifts of this structure are needed to confirm its existence and measure its velocity dispersion. Still, even if each overdensity is confirmed with spectroscopy, they could still be unbound and coupled to the Hubble flow. In this case, they possibly trace a filament in the dark matter density field and may evolve into a large-scale structure at $z=0$, e.g., a supercluster \citep{geller_mapping_1989}.

To help us interpret the $z=2.2$ system as a whole, we searched for analogs in the $1$~Gpc$^3$ GiggleZ cosmological simulation (Poole et al., in prep.; see \$\ref{highz}) by running 2~million random line of sight (each a cylinder with $0.7$ Mpc radius and $50$ Mpc length to match the $2\sigma$ redshift uncertainty, $2\sigma_{z}\approx0.06$, of the overdensities).  If a 3-halo system matching the set of overdensity \mh\ estimates is found, then 98\% percent of the time 2 or 3 of these halos will merge into a $z=0$ cluster with \mh\ $\sim10^{14}-10^{15}$ \msun.  At $z=2.2$ in the simulation, these 2- and 3-halo systems show typical maximum separations of $\sim0.9-1.8$ Mpc 3D from their mean positions.  Although this comparison is limited by the preliminary halo masses of the $z=2.2$ overdensities, these results may suggest there is high probability that two or more of the $z=2.2$ overdensities will merge by $z=0$.

If two or more of the overdensities are currently gravitationally bound, we may be viewing merging sub-clusters that are each in different evolutionary stages, including some whose galaxies are rapidly evolving and only just forming their red galaxy population (e.g., overdensity~C). Perhaps this signifies the system is in a transitional phase between the known $z\ga2$ ``protoclusters'' \citep[e.g.][]{steidel_spectroscopic_2005} with more diffuse distributions of blue galaxies, and the lower-redshift galaxy clusters with prominent red sequences.  Of course, protocluster observations have largely been optically-based, thus deep near-IR observations of such structures and deep optical spectroscopy of our candidate cluster are needed to understand the relationship between the various structures at $z\ga1.5$.\\

The discovery of the $z=2.2$ system demonstrates the powerful combination of near-IR medium-bandwidth filters and deep imaging of the $Z-$FOURGE survey. These systems were undetectable in earlier optical and near-IR catalogs. For example, in the optically-selected ($i<25$) catalog of \citet{ilbert_cosmos_2009}, the overdensity is completely absent in our redshift slice of interest ($z=2.1-2.3$).  The relatively bright optical limit of this catalog means evolved galaxies at these redshifts are missed entirely (only 2 of our red overdensity galaxies have $i_{total}<25$, see Fig.~\ref{fig_cmd}).  Even with the $K$-band selected NMBS catalog of \citet{whitaker_newfirm_2011}, only a weak overdensity is seen in Fig.~\ref{fig_density}, as the majority of the $z=2.2$ overdensity galaxies are too faint to be detected in the 1.3 mag. shallower NMBS.  

The first results of the $Z-$FOURGE survey suggest we are reaching a critical threshold in our ability to study galaxy evolution as a function of local environment at $z\ga1.5$. By combining the spatial distribution with improved redshift information from deep medium-band filters in the near-IR, we can start to correlate the properties of $K_s-$selected galaxies with their environment, something that has so far only been done in detail for Lyman-break galaxies \citep[e.g., ][]{steidel_ly_2000}.
\acknowledgments
We acknowledge the Cook's Branch Conservancy for the gracious hospitality and comfortable surroundings which permitted the discovery of this exciting system.  We are grateful to the \fourstar\ instrumentation team and staff at LCO.  We appreciate the useful comments from the anonymous reviewer. LRS acknowledges funding from a Australian Research Council (ARC) Discovery Program (DP) grant DP1094370 and Access to Major Research Facilities Program which is supported by the Commonwealth of Australia under the International Science Linkages program.  GBP acknowledges support from two ARC DP programs (DP0772084 and DP1093738).  CP, KVT and VT acknowledge support from National Science Foundation grant AST-1009707. Australian access to the Magellan Telescopes was supported through the National Collaborative Research Infrastructure Strategy of the Australian Federal Government.

{\it Facilities:}\facility{Magellan(\fourstar)}.

\bibliographystyle{apj}
\bibliography{ms}

\end{document}